\documentclass[%
 reprint,
superscriptaddress,
 amsmath,amssymb,
 aps,
]{revtex4-2}

\usepackage{graphicx}
\usepackage{dcolumn}
\usepackage{bm}

\begin{document}

\preprint{APS/123-QED}

\title{Gallium nitride phononic integrated circuits for future RF front-ends} 

\author{Mahmut Bicer}
\address{Quantum Engineering Technology Labs and Department of Electrical and Electronic Engineering, University of Bristol,
Woodland Road, Bristol BS8 1UB, United Kingdom}

\author{Stefano Valle}
\address{Quantum Engineering Technology Labs and Department of Electrical and Electronic Engineering, University of Bristol,
Woodland Road, Bristol BS8 1UB, United Kingdom}

\author{Jacob Brown}
\address{Quantum Engineering Technology Labs and Department of Electrical and Electronic Engineering, University of Bristol,
Woodland Road, Bristol BS8 1UB, United Kingdom}

\author{Martin Kuball}
\address{Centre for Device Reliability and Thermography and H. H. Wills Physics Laboratory, University of Bristol,Tyndall Avenue, Bristol BS8 1TL, United Kingdom}

\author{Krishna C. Balram}
\email{krishna.coimbatorebalram@bristol.ac.uk}
\address{Quantum Engineering Technology Labs and Department of Electrical and Electronic Engineering, University of Bristol,
Woodland Road, Bristol BS8 1UB, United Kingdom}

\date{\today}

\begin{abstract}
Achieving monolithic integration of passive acoustic wave devices, in particular RF filters, with active devices such as RF amplifiers and switches, is the optimal solution to meet the challenging communication requirements of mobile devices, especially as we move towards the 6G era. This requires a significant ($\approx100x$) reduction in the size of the RF passives, from $mm^{2}$ footprints in current devices to tens of ${{\mu}m^2}$ in future systems. Applying ideas from integrated photonics, we demonstrate that high frequency ($>$3 GHz) sound can be efficiently guided in ${\mu}m$-scale gallium nitride (GaN) waveguides by exploiting the strong velocity contrast available in the GaN on silicon carbide (SiC) platform. Given the established use of GaN devices in RF amplifiers, our work opens up the possibility of building monolithically integrated RF front-ends in GaN-on-SiC.
\end{abstract}

\maketitle


\section{INTRODUCTION}

\begin{figure*}[!htbp]
    \includegraphics[width =1.0 \linewidth]{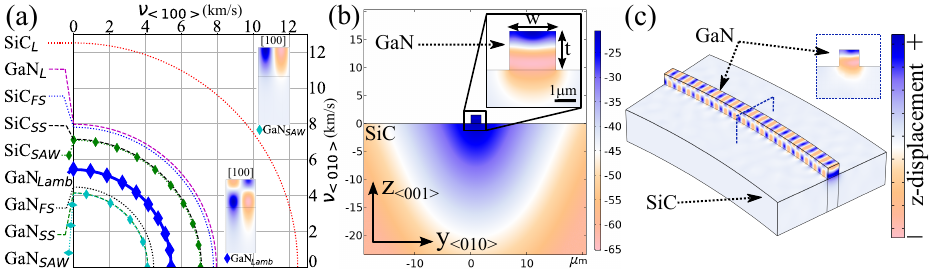}
    \caption{(a) In-plane acoustic velocity for various (bulk and surface) acoustic wave modes in GaN and SiC. The acoustic displacement profiles for the GaN SAW and Lamb wave modes are shown in the inset. (b) FEM simulation ($z$-displacement, log scale) of the propagating Lamb wave mode for a GaN waveguide with dimensions $t=1.5 \ \mu m$ and $w=1.8 \  \mu m$. The confinement of the acoustic energy in the GaN layer and its decay into the substrate can be clearly seen. The inset shows the $z$-displacement of Lamb wave mode in a linear scale. (c) 3D FEM simulation ($z$-displacement) of an acoustic wave mode propagating around a bend.}
    \label{fig:slowness curve and mode profiles}
\end{figure*}

Modern smartphones have expanded the scope and range of functions expected from a mobile device far beyond cellular communications, placing stringent constraints on RF filtering as we move towards the 6G era\cite{lam2016review}. A state-of-the-art smartphone has $>$50 RF filters \cite{ruby2015snapshot} and this number is expected to increase significantly in future generations, pushing the limits of the discrete integration and packaging \cite{balysheva2019saw} approaches employed by current multi-band multi-chip modules. Monolithic integration of acoustic wave filters, with active devices like RF amplifiers and switches, is the only viable long-term solution to this problem\cite{HUSNAHAMZA2020153040,khalil2008gan}.

\par To achieve monolithic integration, two key challenges need to be addressed: the choice of a material platform in which system-level performance, especially insertion loss, comparable to available discrete solutions can be demonstrated; and a reduction in the footprint of acoustic wave devices from $mm^2$ to tens of ${\mu}m^2$ per filter to truly exploit the benefits of dense integration. Current filter sizes at $\approx mm^2$ per filter occupy a footprint much larger than the acoustic wavelength ($\lambda_a \approx1 \ {\mu} m$) in these structures. Unlike in integrated photonics, where wavelength-scale devices like microring resonators and photonic crystal cavities are routinely used, in acoustic devices, the footprint is $\approx10^3\lambda_{a}^2$. This is mainly because current acoustic wave devices, both bulk and surface wave based, employ weak transverse confinement of the acoustic field. By applying ideas from integrated photonics to high-frequency acoustic waves and working in platforms with strong acoustic index contrast, phononic integrated circuits (PnIC) that can confine high-frequency sound to the wavelength scale can be designed and fabricated to dramatically reduce the footprint of acoustic wave devices.

While the analogy between photonics and phononics has been extensively explored\cite{laude2015phononic, safavi2019controlling}, there has been relatively little application of these ideas to the problem of shrinking the footprint of traditional acoustic wave devices, especially at GHz frequencies\cite{liu2017toward, benchabane2019elastic}. Murata's recent IHP-SAW \cite{murata2016} demonstration shows the benefits of acoustic wave confinement. While 1D SAW confinement exhibits a significant improvement in device performance, it does not lead to a reduction in device footprint. By moving towards 2D confinement of sound and building a PnIC platform \cite{siddiqui2018lamb,dahmani2020piezoelectric,mayor2021gigahertz} a significant reduction in the footprint of these micro-acoustic components can be achieved\cite{fu2019phononic}.

In this work, we demonstrate GaN-on-SiC as an ideal platform for monolithic integration of RF passives with active devices and show a proof-of-principle PnIC platform that can efficiently route and manipulate high-frequency ($>$3 GHz) sound waves in ${\mu}m$-scale waveguides and resonators. While the piezoelectric coefficient of GaN is lower than traditional acoustic wave filter materials like aluminum nitride (AlN) and lithium niobate (LN), the GaN-on-SiC platform provides some unique advantages. In particular, the system supports low-loss guided acoustic modes due to the massive acoustic velocity contrast between the GaN device layer (slow) and the SiC substrate (fast). In addition, thin (1-2 ${\mu}m$) GaN-on-SiC films can be grown with sufficiently low dislocation density so that high-performance guided acoustic wave devices can be engineered\cite{valle2019high}. We believe that the overall footprint of RF front-ends can be reduced without affecting the overall system performance, in particular insertion loss, by engineering tight integration between the active and passive components on the same die \cite{ghosh2019acoustoelectric}. While this paper mainly focuses on footprint reduction, in section \ref{outlook}, we also outline some routes to improved system-level performance.

\section{Design \&\ FEM modelling of acoustic wave propagation}

\par GaN-on-SiC provides an almost perfect realization of a slow-on-fast platform as shown in the Fig.\ref{fig:slowness curve and mode profiles}(a), which plots the in-plane velocity (spanned over angles ranging from 0 to $\pi$/2 from [100]) for bulk (longitudinal, slow, and fast shear waves) and surface (SAW) modes in GaN and SiC. The displacement profiles for some representative modes are shown in the inset. The Lamb wave mode, which is the main focus of this work, is clearly separated by $\approx1300 \ m/s$ from the nearest substrate mode (the SAW mode of SiC) throughout the $x-y$ plane, ensuring that the acoustic wave can be routed and guided with minimal substrate leakage in any direction in-plane.   

This is confirmed by Fig.\ref{fig:slowness curve and mode profiles}(b) which shows an FEM simulation of the $z$-displacement (log-scale) for a Lamb wave mode at frequency $\approx3.4$ GHz in a GaN-on-SiC strip waveguide with dimensions $t=1.5 \ \mu m$, $w=1.8 \ \mu m$. The acoustic energy is primarily confined to the GaN layer and decays into the substrate with a decay length $\approx \lambda_{a}$, where $\lambda_a$ is the acoustic wavelength ($1.6\ \mu m$). We chose to work with the Lamb wave mode for two main reasons: the higher acoustic velocity ($\approx5414 \ m/s$) compared to the SAW mode enables more efficient access to higher frequencies (3-6 GHz), and the tighter confinement in the GaN layer (Fig.\ref{fig:slowness curve and mode profiles}(b)) which results in a higher transduction efficiency (and lower insertion loss) when excited by an interdigitated transducer (IDT) \cite{murata2016, valle2019high}. Fig.\ref{fig:slowness curve and mode profiles}(c) shows the propagation of the mode around a bend with radius $R=120\ \mu m$ showing negligible acoustic radiation loss into the substrate.

An interesting observation from Fig.\ref{fig:slowness curve and mode profiles}(a) is that the GaN-on-SiC platform has negligible in-plane velocity anisotropy ($0.26\%$), especially compared to other piezoelectric substrates like LN and GaAs \cite{de2003focusing}. This implies that standard focusing IDTs can be used to efficiently interface RF signals with these ${\mu}m$-scale waveguides \cite{siddiqui2018lamb}, without requiring anisotropy correction to the finger shape\cite{de2003focusing} and any constraints on the focusing IDT aperture. This has implications for the overall achievable efficiency because one of the key challenges with PnICs is the design of efficient transducers that are simultaneously impedance-matched to 50 $\Omega$ and can mode-matched to ${\mu}m$-scale waveguides. The standard focusing IDT \cite{msall2020focusing} is one way to address this problem, and avoiding velocity anisotropy (and the associated shape correction) potentially helps to improve the overall transduction efficiency \cite{balram2021piezoelectric}. 

\begin{figure}[!hbtp]
{\includegraphics[width = 1.0 \columnwidth]{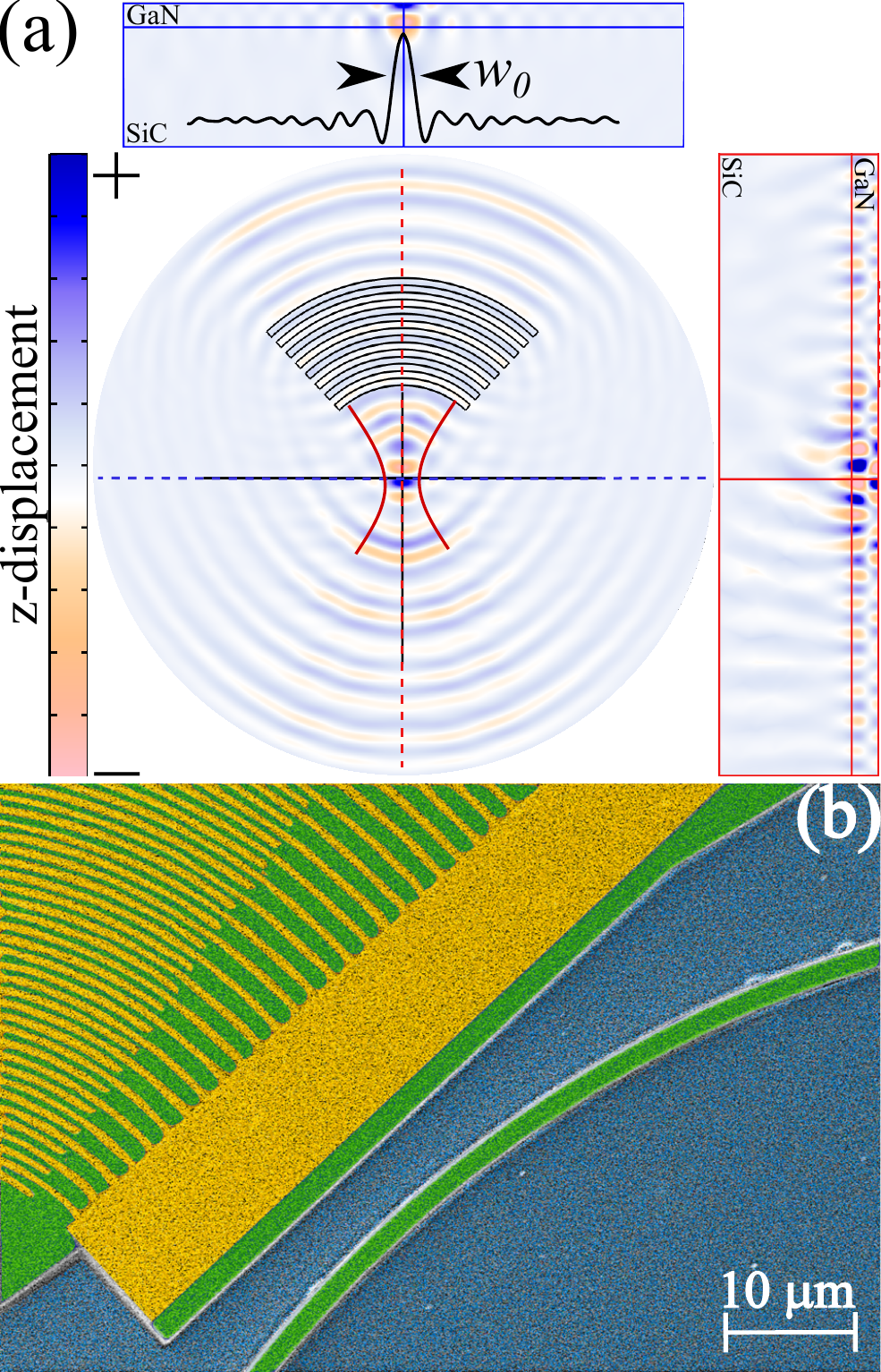}}
  \caption{(a) 3D FEM simulation of the focused acoustic field generated by a curved IDT ($\lambda = 1.6\  \mu m$). The acoustic displacement in the longitudinal and transverse focal sections are shown in the inset. We estimate the beam waist to be $\approx1.67\  {\mu}m$. (b) False-colored SEM image of a focused IDT and a micro-ring resonator, (green = GaN, yellow = gold and blue = SiC substrate)}
  \label{fig:2 Focusing IDT}
\end{figure}

To maximize the coupling of sound (Lamb waves) from a focusing IDT into a GaN strip waveguide, the waveguide width needs to be optimized to match the focusing spot size. This is the acoustic analog of the optical mode-matching problem, although we would like to note here that the acoustic version is complicated by the existence of surface modes at the interface (see section \ref{outlook}). It is easy to see that if the waveguide width is too small compared to the spot size, one intercepts only a fraction of the input power. On the other hand, if the waveguide is too wide, then the input spot excites higher order propagating modes which leads to excess detection loss at the output IDT. To pick the optimal waveguide width, we can refer to Fig.\ref{fig:2 Focusing IDT}, which shows the acoustic field generated by a focusing IDT exciting a Lamb wave mode in GaN-on-SiC with an angular span $1.6$ radians. The longitudinal and transverse cross sections of the field in the focusing plane are also shown in the inset. The focused acoustic field can be well approximated by a Gaussian mode with a focused spot size of $\approx1.67\ {\mu}m$. In our experiments, we chose to work with a slightly larger width of $\approx1.8 \ {\mu}m$ to account for fabrication tolerances. 

\section{Device Characterisation}

To characterize acoustic wave propagation through ${\mu m}$-scale waveguides in GaN-on-SiC PnICs, the structures are designed as emitter-receiver pairs. Fig.\ref{fig:2 Focusing IDT}(b) shows a false-colored SEM image of a fabricated device. The devices are fabricated using a multilayer aligned e-beam lithography process. The dry etching of the GaN is optimized for low sidewall roughness to minimize surface scattering losses. The focusing IDT structures are patterned with electrode period ($\lambda$) 1.6 ${\mu}m$ using Cr/Au electrodes with thickness 5/45 nm respectively. A vector network analyzer is used to launch high-frequency ($\approx3.5$ GHz) Lamb waves using focusing IDT structures (Fig.\ref{fig:2 Focusing IDT}(b)). These Lamb waves are routed through straight waveguides, bends, and microring resonators and detected using a symmetric focused IDT structure which functions as a coherent receiver.

\begin{figure}[!htbp]
{\includegraphics[width = \columnwidth]{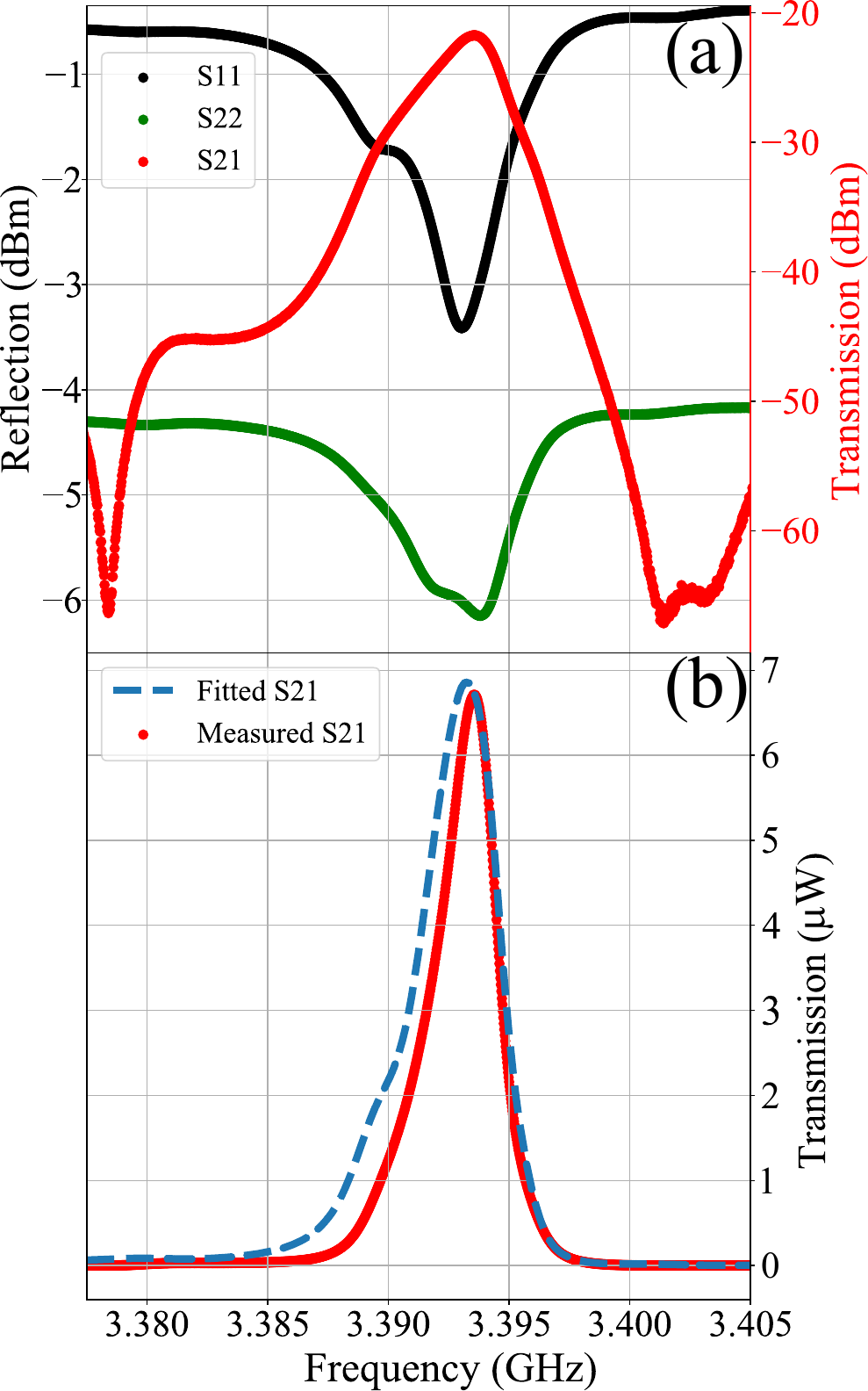}}
  \caption{(a) Measured RF reflection spectra ($S_{11}$ and $S_{22}$) of the emitter and receiver focusing IDTs. The RF transmission ($S_{21}$) spectrum through a 64 ${\mu}m$ long, 1.8 ${\mu}m$ wide waveguide is also plotted (red) with a transmission peak of -21.7 dB.  (b) Linear transmission ($S_{21}$) spectrum and best-fit using a P-matrix model (details in text) to extract the various system parameters.}
  \label{fig:3 straight waveguide}
\end{figure}

Experimental data from a device connecting two focusing IDTs ($\lambda =  1.6 \ {\mu}m$) with a waveguide ($w_{wvg} = 1.8 \ \mu m $, $ \ L_{wvg} = 64 \ {\mu} m$) is shown in Fig.\ref{fig:3 straight waveguide}(a). The RF reflection ($S_{11}$ and $S_{22}$) spectra of the emitter and receiver focusing IDTs are shown in black and green respectively and the transmission ($S_{21}$) spectrum is shown in red. Fig.\ref{fig:3 straight waveguide}(a) clearly shows that efficient energy transfer can only be achieved if the spectra of the emitter and receiver IDTs line up (discussed further below). The peak transmission, which is a measure of the overall insertion loss, achieved in this simple geometry is -21.7 dB. 

To quantify the system parameters and understand the sources of loss inside the circuit, we use a modified P-matrix formulation \cite{datta1986surface} (see Appendix I). There are three main sources of loss for this simple structure consisting of two focusing IDTs connected by a ${\mu}m$-scale waveguide of length $L$. The IDTs are bi-directional and roughly half the acoustic energy (in both the emitter and receiver IDTs) is lost. There is further loss due to reflection at the waveguide-taper interface, where the GaN layer is flared out to accommodate the IDT. Finally, there is the propagation loss in the waveguide itself. Using the IDT's measured $S_{11}$ and $S_{22}$ response as input, we can fit the measured $S_{21}$ spectrum using the P-matrix model. In the model, we assume the structure is fully symmetric (the reflection loss at the entrance and exit of the waveguide are identical) and both IDTs couple half of their power in the forward direction. Given the short waveguide lengths, the waveguide-taper interface dominates the measured transmission loss. By modeling the interface as a semi-transparent mirror (a beam-splitter), the P-matrix fit to the experimental data allows us to extract an interface reflectivity of 17.5$\%$. This implies only 82.5$\%$ of the power excited by the IDT in the forward direction is coupled into the Lamb wave mode of the waveguide. Designing a mode-matched interface is therefore critical to achieving high efficiency and low insertion loss, discussed further in section \ref{outlook}. 

\begin{figure}[!htbp]
{\includegraphics[width = \columnwidth]{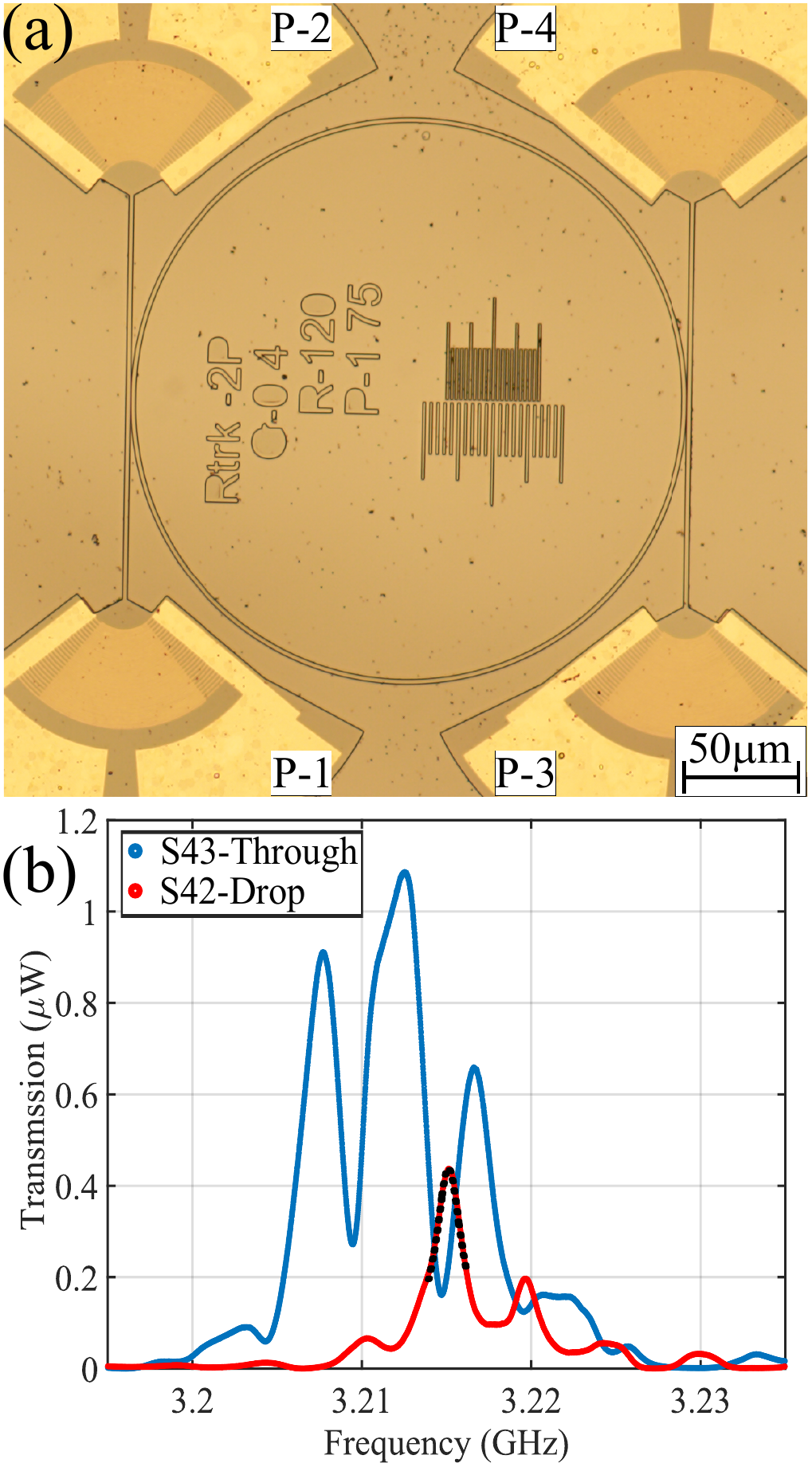}}
  \caption{(a) Microscope image of an acoustic microring resonator (R = 120 ${\mu}m$) with side coupled waveguides (gap = 300 nm) to enable a full 4-port measurement of the device. (b) RF transmission spectra ($S_{21}$) measured in the Through (blue) and Drop (red)  ports, clearly indicating the spectrally aligned dips and peaks corresponding to successive acoustic microring resonator modes. The Lorentzian fit of one of the resonator modes is indicated by the dashed black curve.}
  \label{fig:4_microring_resonator}
\end{figure}

One of the main advantages of being able to route sound effectively around tight bends in a PnIC platform is the prospect of using acoustic microring resonators to store and filter sound on a chip. In analogy with photonics, the prospect of using acoustic whispering gallery modes and total internal reflection of sound for confinement is expected to lead to lower scattering and insertion losses on a chip, in addition to a massive footprint reduction, although this has not yet been experimentally verified.

Fig.\ref{fig:4_microring_resonator}(a) shows a microscope image of a representative acoustic microring resonator with radius 115 ${\mu}$m, coupled to two strip waveguides, which can be used to filter and route sound between the four ports labeled in the figure. A focusing IDT is used at each port to generate and detect the acoustic wave at that port. The ring radius is set by the need to fit more than one ring resonance (determined by its free spectral range (FSR)) within the excitation and detection bandwidth of the IDTs. The GaN-on-SiC platform can in principle support low-loss microrings with radii $<$20 ${\mu}m$  

Fig.\ref{fig:4_microring_resonator}(b) shows the measured linear transmission ($S_{21}$) spectrum in the through (blue) and drop (red) ports, for a device with $\lambda_{a} = 1.75\  \mu m$, $R = 115 \ \mu m$, and $L_{wvg} = 176 \  \mu m$. The gap between the waveguide and the ring is 300 nm, and the overlap between the waveguide and the ring is $2\lambda_{a}$.  In contrast to the straight waveguide case (Fig.\ref{fig:3 straight waveguide}(a)), the transmission spectrum shows clear signatures (dips) corresponding to the successive microring resonances. The corresponding peaks in the drop port which are spectrally aligned with the dips in the through port can also be seen. In addition, the measured FSR of the ring resonances is $\approx5.2$ MHz, which is close to the predicted FSR of $5.16$ MHz, corresponding to a Lamb wave group velocity of $\approx3800$ m/s. By fitting the resonator response with a Lorentzian lineshape (dashed black curve), we extract a quality factor ($Q_a$) of $\approx1700$. In the limit where the dissipation in the ring resonator is dominated by intrinsic losses \cite{chrostowski2015silicon}, we can bound the waveguide propagation loss ($\alpha_a = \frac{2{\pi}{n_g}}{\lambda_{a}Q_{a}}$) to $\alpha_{a} \approx13.6$ dB/mm, using a group index ($n_g$) of 1.42. To put this number into context, in our previous result on observing trapped guided mode resonances in this platform \cite{valle2019high}, we observed $Q_{a} \approx2000$ for unconfined acoustic modes.

\section{OUTLOOK and future prospects for GaN phononic integrated circuits}\label{outlook}

While in this work, we have demonstrated a proof-of-principle GaN-based PnIC platform that can enable significant footprint reduction for passive acoustic wave devices, the device performance needs to be improved if they are to displace existing solutions. The key figure of merit to target is the insertion loss, as shown by the peak transmission ($S_{21}$) in Fig.\ref{fig:3 straight waveguide}(a). For emitter-receiver IDTs connected by a straight waveguide, our peak transmission is $\approx-21.7$ dB. For state-of-the-art mobile systems, this insertion loss needs to be $<$2 dB to benefit from the massive footprint reduction. In particular, we would like to achieve similar insertion-loss performance compared to traditional acoustic wave devices in these highly reduced form-factors. 

To overcome this insertion loss, it is important to understand the various contributing effects to the -21.7 dB value and look at each in isolation. As discussed above, bi-directional IDTs such as the ones used here give a 6 dB loss, which can be addressed by unidirectional focusing IDTs\cite{ekstrom2017surface}. A further 8 dB loss can be attributed to the emitter and receiver IDTs not being impedance matched, leading to less than half the input RF power being coupled into (and back out from) acoustic waves. IDTs that are better matched to 50 $\Omega$ should address this loss problem. Based on the P-matrix fits, we estimate a further 2.2 dB loss due to mode matching at the input and exit of the waveguide. By designing adiabatic mode transformers similar to integrated photonics, the reflection loss at the waveguide input and exit can be significantly reduced. Finally, we have $\approx1.7$ dB of propagation loss, based on the separation between the emitter and receiver IDTs.  

Summing up these loss contributions, we predict an overall loss of $\approx17.9$ dB, which leaves us with an excess loss of 3.8 dB in our experiments. We attribute this mainly to roughness-induced scattering and mode conversion in our devices.  By implementing the suggested design and fabrication improvements, mainly improved GaN dry etching and surface cleaning, we believe the insertion loss in these highly confined geometries can be pushed to the $<$2 dB limit in the near-term, making this platform and architecture technologically viable for future RF front-ends. 

\begin{figure}[!htbp]
{\includegraphics[width = \columnwidth]{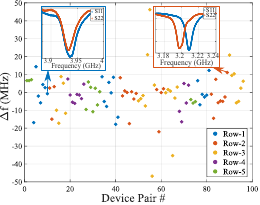}}
  \caption{Plot showing the mismatch in central frequency (${\Delta}f$) for 100 nominally identical emitter-receiver focusing IDT pairs characterized in our experiments. The inset of the plot shows the spectral response of the two IDTs in the pair in an exemplary matched response and mismatched response case.}
  \label{fig:5_idt_mismatch}
\end{figure}

There is an additional source of loss that occurs due to nanofabrication errors. Given the emitter and receiver IDTs are relatively narrowband structures, lining up their spectral responses (Fig.\ref{fig:3 straight waveguide}(a)) is critical to achieving high transmission efficiency. The data reported in Figs.\ref{fig:3 straight waveguide} and \ref{fig:4_microring_resonator} are for the best-performing devices in our measurements. In general, we observe a significant spread in the response of nominally identical devices due to surface charging during e-beam lithography. This data is plotted in Fig.\ref{fig:5_idt_mismatch} where the difference in central frequency (${\Delta}f$) for 100 nominally identical focusing IDT emitter-receiver pairs is shown. As can be seen, there is a significant spread in the data of the order of $\approx10$ MHz, which is close to the IDT bandwidth ($\approx20$ MHz). The inset to the plot shows the measured spectral responses in representative matched and mismatched cases. In our experiments, we find that matched $S_{11}$ responses of the emitter and receiver IDTs are critical to observing high $S_{21}$ and mismatched IDTs can have spectral responses that are suppressed by as much as 30 dB. This dataset points to the critical importance of controlling nanofabrication to ensure low insertion loss in fabricated devices.

\section{conclusions}

In this work, we have shown a proof-of-principle GaN PnIC platform demonstrating efficient routing of high-frequency sound in ${\mu}m$-scale waveguide and resonators. Our work opens up the path towards developing monolithically integrated RF front-ends in GaN by interfacing GaN HEMT amplifiers with these confined acoustic wave devices.

\section{Acknowledgements}

We would like to thank the Engineering and Physical Sciences Research Council (GLIMMER, EP/V005286/1 and QuPIC, EP/N015126/1) and European Research Council (ERC-StG, SBS3-5, 758843) for funding support. JB would like to thank the Engineering and Physical Sciences Research Council for a DTP award (EP/T517872/1). The authors would like to thank Manikant Singh, Martin Cryan, Bruce Drinkwater and John Haine for useful discussions and suggestions.

\section{Appendix}
\subsubsection{P-Matrix Formulation}
\par To further analyze and improve the system, P-matrix formulation is employed\cite{morgan2010surface}. The main idea of this model is to create an analytical formula to extract the reflectivity parameter at the entrance of the waveguide, through a fitting algorithm. In this model, each section of this system is expressed in a P-matrix form. These sections are taper ($P_t$), waveguide ($P_{wg}$), and interface between taper and waveguide ($P_i$). 
IDTs are modelled as a black box that converts electrical power to displacement amplitude and vice versa, therefore they are not expressed in P-matrix. This effectively makes the P-matrix formulation equivalent to an acoustic transfer-matrix. The amplitude of the acoustic wave ($A$) launched by IDT is $A = \sqrt{2P_s}$, where $P_s$ is the power transmitted by IDT in one direction, and it is equal to $P_s = (1-|S_{11}|^2)/2$.

The taper and waveguide sections are assumed as lossless, given the short propagation lengths. Therefore they can be represented as a delay line in the formula, according to:

\begin{equation}
P_t =
\begin{bmatrix}
p_{11}&p_{12}\\
p_{21}&p_{22}
\end{bmatrix}
=\begin{bmatrix}
0&e^{-i\beta l_t}\\
e^{-i\beta l_t}&0
\end{bmatrix}\end{equation}
\begin{equation}
P_{wg} = 
\begin{bmatrix}
p_{11}&p_{12}\\
p_{21}&p_{22}
\end{bmatrix}
=\begin{bmatrix}
0&e^{-i\beta l_{wg}}\\
e^{-i\beta l_{wg}}&0
\end{bmatrix}\end{equation}
On the other hand, the interface between the taper and the mirror is modeled as a semi-transparent mirror. If we assume the mirror to have ‘$r$’ reflection coefficient and '$t$' transmission coefficient for both reference planes, then we can relate P-matrix elements as $p_{11}=p_{22}=-r$,  and  $p_{12}=p_{21}=t$ . For lossless reciprocal 2-port system, P-matrix elements need to satisfy following condition; 
\begin{equation}
|p_{11}|^2 + |p_{21}|^2 = 1 , 
|p_{22}|^2 + |p_{12}|^2=1
\end{equation}
If $r$ and $t$ coefficients are substituted, we find the $ t=\sqrt{(1-r^2)}$ relation. Lastly, for unitary of the system, following condition need to be met;
\begin{equation}
p^{*}_{11}p_{12} + p^{*}_{21}p_{22}=0
\end{equation}
\begin{equation}
p_{12} = - p^{*}_{21}(p_{22}/p^{*}_{11})
\end{equation}
For the symmetry of the reference plane, $p_{22}/p^*_{11}$ is equal to unity. Therefore,  $p_{12}$ and $p_{21}$ can only be imaginary, $p_{12}=p_{21}=jt$. From these equations and relations, $P_i$ can be written as:
\begin{equation}
P_i =
\begin{bmatrix}
p_{11}&p_{12}\\
p_{21}&p_{22}
\end{bmatrix}
=\begin{bmatrix}
-r&jt\\
jt&-r
\end{bmatrix}\end{equation}

Final P-matrix equation of the system is obtained by multiplication P-matrices of each section (eqn.7)
\begin{equation}
P = P_t P_i P_{wg} P_i P_t  
\end{equation}

\bibliography{GaN_Main_aip}

\end{document}